# Close supermassive binary black holes

**It has been proposed that when the peaks of the broad emission lines in active galactic nuclei (AGNs) are significantly blueshifted or redshifted from the systemic velocity of the host galaxy, this could be a consequence of orbital motion of a supermassive blackhole binary (SMB)[1]. The AGN J1536+0441 (=SDSS J153636.22+044127.0) has recently been proposed as an example of this phenomenon[2]. It is proposed here instead that 1536+044 is an example of line emission from a disc. If this is correct, the lack of clear optical spectral evidence for close SMBs is significant and argues either that the merging of close SMBs is much faster than has generally been hitherto thought, or if the approach is slow, that when the separation of the binary is comparable to the size of the torus and broad-line region, the feeding of the black holes is disrupted.**

Galaxies grow through mergers, and since all massive galaxies contain supermassive black holes, the formation of SMBs will be common[1,3,4]. Close SMBs, with orbital velocities, $v_{orb} \sim 0.01c$, are expected to last long enough to be observed[1,3]. It was proposed[1] that such close binaries could be detected from velocity shifts of their broad-line region (BLR) line profiles, and pointed out that velocity shifts of the peaks of low-ionization broad lines are common[1,4,5].

There are two testable predictions of this model: firstly, the radial velocities of the peaks in the line profiles will shift on the orbital timescale of the SMB[6], and secondly, since all AGNs vary, if there are two separate BLRs, the line fluxes of the two peaks will vary *independently*[7]. The prototypical displaced BLR peak AGN 3C 390.3 fails both these tests. The radial velocity signature of orbital motion has been detected[6,8] but the velocity changes are incompatible with an SMB and are instead consistent with orbital motion of features in a non-azimuthally-symmetric disc. The peaks in 3C 390.3 initially seemed to be varying independently on timescales longer than the light-crossing timescale[7], but an SMB is strongly ruled out by better-sampled monitoring[9,10] which shows that on a light-crossing time the peaks vary *simultaneously* as expected for a disc. The longer timescale profile changes in these AGNs are consistent with orbital motion of clumps in a disc[11]. The profiles are consistent with theoretical line emission profiles expected from discs[12], but the discs are generally not azimuthally symmetric and it is common for one peak to be significantly stronger than the other[1,4,5,12].

The Balmer lines of J1536+044 have a strong blue-shifted peak[2]. The Hβ profile is shown in Fig. 1. Boroson & Lauer[2,13] interpret J1536+044 as a SMB. I argue here, however, that just as the Balmer line profiles in previous SMB candidates have been shown to be due to disc emission, so too the Balmer line profile of J1536+0441 probably arises from disc emission. This has also been independently proposed by Chornock et al.[14,15].

Although J1536+044 represents an extremum among AGNs selected as "quasars" by the SDSS, its line profiles are not unique among AGNs in general. The AGN 0945+076, for example, has shown a nearly identical Hβ profile, as can readily be seen by comparing

Fig. 1 with the Hβ profile of 0945+076 in ref. 1. To qualitatively illustrate the non-uniqueness of J1536+044, Fig. 1 includes part of the scaled Hβ profile of the well-known disc emitter Arp 102B. Fitting disc models to theoretical double-peaked profiles invariably shows that there is an extra component of gas at the systemic velocity (see, for example, any of the fits in ref. 12), so the spectral region with a width corresponding to that of the high-velocity wings of the [O III] λ5007 line has been excluded around rest-frame Hβ. The width of the Arp 102B spectrum has been reduced by 34% to make the velocity of the blueshifted peak in J1546+0441 match that of the blueshifted peak in the mean Arp 102B spectrum. This is well within the range of line widths seen among disk-like emitters. The uncertainty in the width scaling is about ± 5%. The flux has been scaled by minimizing residuals away from the peaks of the broad lines and away from the contaminating [O III] λ5007 lines. The profiles of broad disc-like lines vary strongly, so we do not expect a perfect match between any two objects a given time. The peaks of Arp 102B are particularly variable (see ref. 11). Fig. 1 shows that the differences in the peak fluxes are only about one standard deviation from the scaled Arp 102B mean profile. More recent spectra of J1546+0441[13,15] (especially of Hα) show better agreement in the red peak.

Clearly, as already pointed out by Boroson & Lauer[2,13], the most rigorous test of the competing hypotheses is line profile variability. If this verifies that the Hβ profile of J1536+044 is the result of normal disc emission, it has significant implications for the evolution of SMBs since J1536+0441 is the only candidate so far for a sub-parsec supermassive black hole binary out of ~17,500 AGNs with $z < 0.70$ in the SDSS[2,13]. Because SMB formation must be common, the absence of clear evidence for close SMBs in AGNs needs to be explained. It suggests either that the lifetime of close SMBs is considerable shorter than originally thought or that they are long-lived and have their feeding interrupted so that activity is greatly reduced.

C. Martin Gaskell
*Astronomy Department, University of Texas, Austin, TX 78712, USA*

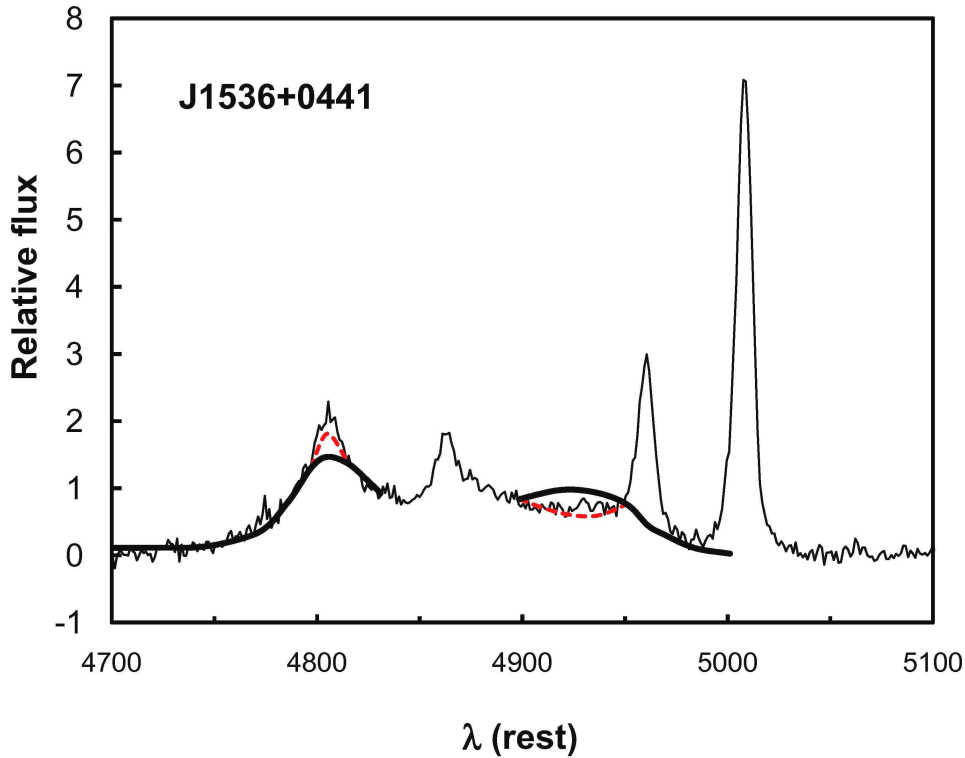

**Figure 1** – Comparison of the continuum-subtracted SDSS spectrum of the Hβ region of J1536+0441 (thin black line) with the mean Hα profile of Arp 102B from 1992-1996 (smooth thick black line). The latter is taken from the mean spectrum shown in Fig. 3 of Sergeev, Pronik, & Sergeeva[11] and scaled as described the text. The dotted red lines illustrate the effects on the peaks in the Arp 102B profile if they are changed by 1σ (based on the rms spectrum in Fig. 4b of ref. 11). The contribution of lower-velocity gas at the systemic velocity in Arp 102B has been omitted.

**Acknowledgements** This research has been supported in part by the U. S. National Science Foundation through grant AST 08-03883. This paper has made use of data from the Sloan Digital Sky Survey (SDSS; http://www.sdss.org/) which is funded by the Alfred P. Sloan Foundation, the U.S. National Science Foundation, the U.S. Department of Energy, NASA, the Japanese Monbukagakusho, the Max Planck Society, the Higher Education Funding Council for the United Kingdom, and participating institutions. The author is grateful to John Kormendy, Miloš Milosavljević, Greg Shields, Remco van den Bosch, and Bev Wills for useful discussions and encouragement.